\documentclass[twocolumn,showpacs,aps,prl,superscriptaddress]{revtex4}

\usepackage{graphicx}
\usepackage{dcolumn}
\usepackage{amsmath}
\usepackage{epsfig}

\input babarsym

\newcommand{\BABARPubYear}    {04}
\newcommand{\BABARPubNumber}  {10}

\newcommand{\SLACPubNumber} {10395}
\newcommand{\LANLNumber} {0404006}

\def\btosee      {\ensuremath{b  \to s\: e^+ e^-}}
\def\btosll      {\ensuremath{b  \to s\: \ell^+ \ell^-}}
\def\BtoXsll     {\ensuremath{\B \to X_s\: \ell^+ \ell^-}}
\def\BtoXsee     {\ensuremath{\B \to X_s\: e^+ e^-}}
\def\BtoXsmumu   {\ensuremath{\B \to X_s\: \mu^+ \mu^-}}
\def\BtoXsemu    {\ensuremath{\B \to X_s\: e^\pm \mu^\mp}}

\def\BtoKll      {\ensuremath{\B \to K \ell^+ \ell^-}}
\def\BtoKstarll  {\ensuremath{\B \to K^\ast \ell^+ \ell^-}}

\def\mee         {\ensuremath{m_{e^+ e^-}}}
\def\mmumu       {\ensuremath{m_{\mu^+ \mu^-}}}
\def\mll         {\ensuremath{m_{\ell^+ \ell^-}}}
\def\mXs         {\ensuremath{m_{X_s}}}

\def\figurebox#1#2#3{%
    \def\arg{#3}%
    \ifx\arg\empty
    {\hfill\vbox{\hsize#2\hrule\hbox to #2{\vrule\hfill\vbox to #1{\hsize#2\vfill}\vrule}\hrule}\hfill}%
    \else
    {\hfill\epsfbox{#3}\hfill}%
    \fi}

\begin{document}

\preprint{\babar-PUB-\BABARPubYear/\BABARPubNumber} 
\preprint{SLAC-PUB-\SLACPubNumber} 

\begin{flushleft}
\babar-PUB-\BABARPubYear/\BABARPubNumber\\
SLAC-PUB-\SLACPubNumber\\
hep-ex/\LANLNumber\\[10mm]
\end{flushleft}

\title{
{\large \bf
 Measurement of the \BtoXsll\ Branching Fraction \\
 with a Sum over Exclusive Modes }
}

%
\author{B.~Aubert}
\author{R.~Barate}
\author{D.~Boutigny}
\author{F.~Couderc}
\author{J.-M.~Gaillard}
\author{A.~Hicheur}
\author{Y.~Karyotakis}
\author{J.~P.~Lees}
\author{V.~Tisserand}
\author{A.~Zghiche}
\affiliation{Laboratoire de Physique des Particules, F-74941 Annecy-le-Vieux, France }
% ...
\collaboration{The \babar\ Collaboration}
\noaffiliation

\date{\today}

\begin{abstract}

We measure the branching fraction for the
flavor-changing neutral-current process \BtoXsll\ with
a sample of $89 \times 10^{6}$~$\Upsilon(4S) \to \BB$ events 
recorded with the \babar\ detector at the \pep2\ $e^{+}e^{-}$ storage ring.
The final state is reconstructed from \epem\ or \mumu\ pairs
and a hadronic system $X_s$ consisting of one $K^{\pm}$ or $K^{0}_{s}$
and up to two pions, with at most one $\pi^{0}$. 
We observe a signal of $40 \pm 10({\rm stat}) \pm 2({\rm syst})$ events
and extract the inclusive branching fraction
${\cal B}(\BtoXsll) = 
 \left(5.6 \pm 1.5({\rm stat}) \pm 0.6({\rm exp~syst})
 \pm 1.1({\rm model~syst})\right) \times 10^{-6}$
for $\mll > 0.2$~\gevcc.

\end{abstract}

\pacs{13.20.He, 12.15.Ji, 11.30.Er}

\maketitle


The rare decay \BtoXsll, which proceeds through the
$b \to s \ell^{+} \ell^{-}$ transition, is interesting
because the study of its rate and charge asymmetry
could lead to indirect observation of physics beyond
the Standard Model (SM).
This transition is forbidden at lowest order in the SM
but is allowed at higher order via electroweak penguin and $W$-box
diagrams.
This implies that non-SM physics in these loops would
contribute at the same order as the SM~\cite{Ali02,Hurth03}.
Recent SM calculations of the inclusive branching fractions 
predict
${\cal B}(\BtoXsee) = (6.9 \pm 1.0) \times 10^{-6}$
[$(4.2 \pm 0.7) \times 10^{-6}$ for $\mee > 0.2$ GeV/$c^2$]
and
${\cal B}(\BtoXsmumu) = (4.2 \pm 0.7) \times 10^{-6}$~\cite{Ali02,Ali_ICHEP02}.
Although the branching fraction measurement for inclusive decays
is more challenging than for exclusive decays,
it is motivated by smaller theoretical uncertainties.
Exclusive $B \to K^{(\ast)} \ell^{+} \ell^{-}$ ($\ell = e, \mu$)
decays have been observed by Belle and \babar~\cite{Belle_Kll,BaBar_Kll}.
Belle has also reported a measurement of
the inclusive \BtoXsll\ branching fraction~\cite{Belle_sll}.

The data sample used in this analysis was collected
with the \babar\ detector~\cite{BaBar}
at the \pep2\ asymmetric-energy $e^{+}e^{-}$ storage ring
at the Stanford Linear Accelerator Center.
The sample consists of $89 \times 10^{6}$ \BB\ events recorded
at the $\Upsilon(4S)$, corresponding to
an integrated luminosity of $81.9$~\invfb.

We study the \BtoXsll\ process by
fully reconstructing a subset of all possible final states.
The reconstructed hadronic systems $X_s$ consist of
one $K^{\pm}$ or $K^{0}_{s}$ 
and up to two pions, with at most one $\pi^{0}$. 
This approach
allows approximately half of the total inclusive rate to be reconstructed.
If the fraction of modes containing a \KL\ 
is assumed equal to that containing a \KS,
the missing states represent $\sim$30\% of the total rate.
To compute the inclusive branching fraction,
we account for missing modes and selection efficiencies
using a \BtoXsll\ decay model constructed as follows.
For $\mXs < 1.1$~\gevcc, 
exclusive $B \to K^{(*)} \ell^{+} \ell^{-}$ decays are generated 
with a $b \to s \ell^+ \ell^-$ decay model according to~\cite{Ali02,Ali00}.
The remaining decays, for $\mXs > 1.1$~\gevcc, 
are generated according to a quark-level calculation~\cite{Ali02,Kruger96}
and the $b$-quark Fermi motion model of~\cite{Ali79}. 
JETSET~\cite{JETSET} is then used to hadronize 
the system consisting of a strange quark and a spectator quark.


The full reconstruction method
exploits the strong kinematic discrimination provided by 
$m_{ES} = \sqrt{E_{beam}^2 - \vec{p}_{B}^{\,2}}$
and $\Delta E = E_{B} - E_{beam}$,
where $E_{beam}$ is the beam energy
and $E_{B}$ ($\vec{p}_{B}$) is the 
reconstructed $B$-meson energy (three-momentum).
These quantities are evaluated in the $e^{+}e^{-}$ center-of-mass (CM) frame. 
 

Clean identification of the $B$ decay products
is important for minimizing the backgrounds.
Electron candidates are required to have a
laboratory-frame momentum $p_\ell > 0.5$~\gevc\
and are identified with measurements from
the tracking systems and the electromagnetic calorimeter.
Bremsstrahlung photons are recovered by
combining an electron with up to three photons
within a small angular region around the electron
direction~\cite{bremrecovery}.
Muon candidates are required to have
$p_\ell > 1.0$~\gevc\ and
are identified as penetrating charged particles in the instrumented
flux return.
Charged kaon identification relies on measurements performed
with the Cherenkov ring-imaging detector and the tracking systems.
The $K^{0}_{s}$ candidates are required to have
$|m_{\pi^{+}\pi^{-}} - m_{K^{0}_{s}}| < 11.2$~\mevcc,
a decay length greater than $2$~mm, and $\cos\delta > 0.99$,
where $\delta$ is the angle between
the $K^{0}_{s}$ momentum vector and a line that connects the primary vertex 
with the $K^{0}_{s}$ vertex.
Charged tracks that do not satisfy tight $e^\pm$ or $K^\pm$
identification are considered to be pions.
Neutral pions are required to have a laboratory-frame energy greater
than $400$~\mev, 
a photon daughter energy greater than $50$~\mev,
and $|m_{\gamma \gamma} - m_{\pi^{0}}| < 10$~\mevcc.

The $B$ candidates are reconstructed by first selecting
the $e^{+} e^{-}$ or $\mu^{+} \mu^{-}$ pair with the largest value of 
$|p_{\ell^{+}}| + |p_{\ell^{-}}|$.
Then, \BtoXsll\ candidates are formed 
by adding any of the following hadronic topologies:
$K^+$, $K^+ \pi^0$, $K^+ \pi^-$, $K^+ \pi^- \pi^0$, $K^+ \pi^- \pi^+$,
$\KS$, $\KS \pi^0$, $\KS \pi^+$, $\KS \pi^+ \pi^0$, and $\KS \pi^+ \pi^-$.
(We name one member of a pair of charge-conjugate states to refer
to both, unless we specify otherwise.)
A limit is imposed on the number of pions because
the expected signal-to-background ratio
drops significantly with increasing multiplicity.


With loose selection criteria,
the average number of $B$ candidates per event
is five in the signal simulation.
A likelihood function for the signal is constructed based on the simulated 
distributions of $\Delta E$, $\log(P_{B vtx})$, and $\cos\theta_{B}$,
where $P_{B vtx}$ is the fit probability for the
$B$ vertex constructed from charged daughter particles and
$\theta_{B}$ is the angle between
$\vec{p}_{B}$
and the $e^-$ beam axis.
We select the candidate with the largest signal likelihood
before further cuts are applied.
This approach entails a loss of 8\% in signal efficiency
but reduces the overall background by 34\% in the final sample.


Combinatorial backgrounds from nonsignal \BB\ and
continuum $e^{+}e^{-} \to q\bar{q}$ (with $q = u,d,s,c$) events
are reduced by requiring
$\mXs < 1.8$~\gevcc,
$5.20 < m_{ES} < 5.29$~\gevcc,
and $-0.2 < \Delta E < 0.1$~\gev. 
We impose a limit on \mXs\ because the
signal-to-background ratio drops as \mXs\ increases.

The dominant background producing a peak in $m_{ES}$ at the $B$ mass
originates from
$B \to J/\psi X$, $\psi(2S) X$ decays with
$J/\psi \to \ell^{+} \ell^{-}$ and $\psi(2S) \to \ell^+ \ell^-$.
This charmonium background is suppressed by
vetoing $B$ candidates with dilepton mass in the ranges
$2.70 < \mee < 3.25$,
$2.80 < \mmumu < 3.20$,
$3.45 < \mee < 3.80$, and
$3.55 < \mmumu < 3.80$~\gevcc.
In the electron channel, the veto is applied
before and after bremsstrahlung recovery to allow
for imperfect recovery.
The potential peaking background from $B \to X_{s} \gamma$ decays
with conversion of the photon into an $e^{+} e^{-}$ pair in
the detector material is
removed by requiring $\mee > 0.2$~\gevcc. 
This cut also serves to reduce contributions
from \btosee\ transitions in which
the $e^{+} e^{-}$ pair originates from a photon.

The final suppression of the combinatorial background is
achieved with a likelihood based on nine variables:
(i) $\Delta E$, (ii) $\Delta E^{ROE}$, (iii) $m^{ROE}_{ES}$,
where $ROE$ refers to the rest of the event
(all charged tracks and photon candidates not included in the $B$ candidate),
(iv) the separation $\Delta z$ between the two leptons
along the beam direction measured at their points of closest
approach to the beam axis,
(v) $\log(P_{B vtx})$, (vi) $\cos\theta_{miss}$, where $\theta_{miss}$ 
is the angle between the missing momentum vector for
the whole event and the $z$~axis in the CM frame,
(vii) $\cos\theta_{B}$,
(viii) $|\cos\theta_{T}|$, where $\theta_{T}$ is the angle between
the thrust axes of the $B$~candidate and the ROE
in the CM frame, and (ix) the ratio $R_{2}$ of the second and zeroth-order
Fox-Wolfram moments~\cite{FoxWolf}.
The variables $\Delta E$, $\Delta E^{ROE}$, and $m^{ROE}_{ES}$
are most effective at rejecting \BB\ background, especially
for events with two semileptonic decays that have large missing energy. 
The event-shape variables $|\cos\theta_{T}|$ and $R_{2}$ are most effective
at suppressing continuum events.
A likelihood value is computed as the product of nine independent probability
density functions (PDF) for the signal, \BB, and continuum background
components.
Using simulated \BtoXsll\ decays, we choose cuts on the ratio
${\cal L}_{R} = {\cal L}^{signal} / 
 ({\cal L}^{signal} + {\cal L}^{\BB} +  {\cal L}^{cont})$
to maximize the statistical significance of the signal.
This optimization is performed separately for 
electron and muon channels in the
regions $\mXs < 0.6$, 
$0.6 < \mXs < 1.1$, and 
$1.1 < \mXs < 1.8$~\gevcc, 
and results in
progressively harder ${\cal L}_R$ cuts for increasing \mXs.

After applying all selection criteria, we obtain a sample of
349 (222) events in the electron (muon) channel.
According to the simulation, the remaining background
consists mostly of \BB\ events.

We perform an extended unbinned maximum-likelihood fit
to the $m_{ES}$ distribution 
to extract the signal yield $N_{sig}$ as well as
the combinatorial background shape and yield.
The likelihood function consists of four components:
signal, charmonium peaking background,
hadronic peaking background (see below), 
and combinatorial background.

The shapes of the signal and charmonium peaking background are
described by the same Gaussian PDF, with a width of 2.80, 2.61,
and 2.74 \mevcc\ in the electron, muon, and electron+muon channels,
respectively.
The Gaussian mean and width are determined
from signal-like $B \to J/\psi X$, $\psi(2S) X$ decay candidates
satisfying all selection criteria
but failing the charmonium veto (charmonium-veto sample).
The charmonium peaking background is estimated from the simulation
to contribute $0.4 \pm 0.2$ ($1.4 \pm 0.5$) events
in the electron (muon) channel.
The size and shape of the hadronic peaking background component arising 
from $B \to D^{(*)} n \pi$~$(n > 0)$ decays with misidentification 
of two charged pions as leptons,
are derived directly from data by performing the analysis without the lepton 
identification requirements.
Taking the $\pi \to \ell$ misidentification rates into
account, we estimate the hadronic peaking background 
to be $<0.1$ and $2.4 \pm 0.8$ events in the electron
and muon channels, respectively.
The error on the misidentification rates derived from data control
samples dominates the uncertainty.
The PDF for the combinatorial background arising from
continuum and \BB\ events is an ARGUS function~\cite{ARGUS} with an
endpoint equal to $E_{beam}$.

\begin{figure}[tb]
\begin{center}
\includegraphics[width=8.7cm]{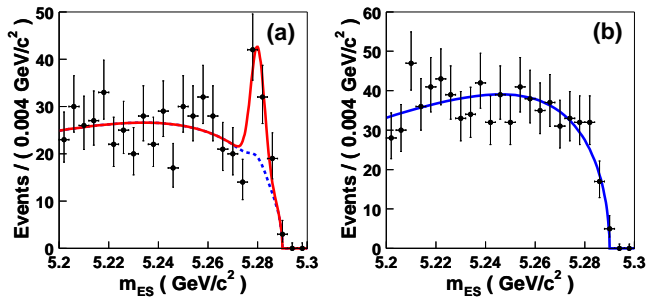}
\caption{Distributions of $m_{ES}$ for selected
(a) \BtoXsll\ ($\ell = e, \mu$) and (b) \BtoXsemu\ candidates.
The solid lines represent the result of the fits
and the dashed line represents the background component
under the signal peak.}
\label{fig:fitfig}
\end{center}
\end{figure}

The fit results are presented in Fig.~\ref{fig:fitfig}(a)
and Table~\ref{tab_results}.
The statistical significance is
${\cal S} = \sqrt{2 \ln({\cal L}_{max} / {\cal L}^0_{max})}$,
where ${\cal L}_{max}$ is the maximum likelihood for the
fit and ${\cal L}^0_{max}$ is that for a
fit with signal yield fixed to zero.
The \BtoXsll\ signal yield
is obtained from a fit to the combined electron and muon data.
Figure~\ref{fig:fitfig}(b) shows the $m_{ES}$ distribution
of \BtoXsemu\ candidates selected using the nominal criteria but
requiring leptons of different flavor. 
An ARGUS function fits this distribution well,
consistent with that sample being pure combinatorial background.

\begin{table}
 \caption{
  Signal yield, significance, efficiency, and branching fraction
  in the electron and muon channels,
  as well as for the electron$+$muon average (first, second,
  and third rows, respectively). 
  For the signal yield,
  the first error is statistical and the second error is systematic.
  For the signal efficiency, the first error corresponds
  to the experimental systematic uncertainty arising from detector modeling,
  hadronization, and Monte Carlo statistics,
  whereas the second error corresponds to
  the uncertainties in the signal model.
  For the branching fraction, the
  errors correspond to statistical, experimental systematic,
  and signal model systematic uncertainties.}
 \begin{center}
 \begin{tabular}{cccc}
  \hline \hline 
  
   $N_{sig}$ & ${\cal S}$ & $\epsilon$ (\%) & ${\cal B}~(10^{-6})$ \\
  \hline\hline
      $29.2 \pm ~8.3 \pm 1.3$ & 4.0
    & $2.74 \pm 0.27 \pm 0.49$
    & $6.0 \pm 1.7 \pm 0.7 \pm 1.1$ \\
      $11.2 \pm ~6.2 \pm 0.9$ & 2.0
    & $1.26 \pm 0.12 \pm 0.25$
    & $5.0 \pm 2.8 \pm 0.6 \pm 1.0$ \\
      $40.1 \pm 10.4 \pm 1.7$ & 4.3
    & $2.00 \pm 0.19 \pm 0.37$
    & $5.6 \pm 1.5 \pm 0.6 \pm 1.1$ \\
  \hline
 \end{tabular}
  \label{tab_results}
 \end{center}
\end{table}

\begin{figure}[tb]
\begin{center}
\includegraphics[width=8.7cm]{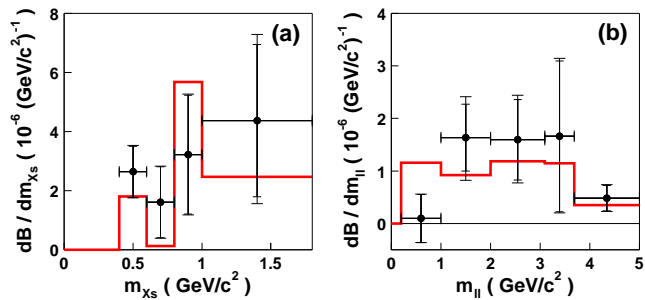}
\caption{Differential branching fraction as a function
 of (a) hadronic mass and (b) dilepton mass,
 averaged over electron and muon channels
 for data (points) and signal Monte Carlo (histogram).
 The outer (inner) error bars correspond to the total (statistical)
 uncertainties.
}
\label{fig:mXs}
\end{center}
\end{figure}

The branching fraction is
${\cal B} = { N_{sig} }/({ 2 N_{\BB}~ \epsilon })$,
where $N_{\BB} = (88.9 \pm 1.0) \times 10^{6}$ is the number
of \BB\ pairs and $\epsilon$ is the signal efficiency.
The yield $N_{sig}$
includes a contribution from
events containing a true \BtoXsll\ decay
but having a selected $B$ candidate with
incomplete or wrong decay products
(cross-feed events).
We estimate the cross-feed contribution
to the signal yield to be $1.5 \pm 1.5$ ($0.6 \pm 0.6$)
events in the electron (muon) channel
by including in the fit a component whose shape
and normalization are taken from simulation.
The corresponding uncertainties are estimated
with an ensemble of simulated data-sized experiments.
The signal efficiency is adjusted to reflect this
contribution.
Figure~\ref{fig:mXs} shows the \BtoXsll\ differential branching
fraction as a function of \mXs\ and \mll, obtained by applying
the nominal likelihood fit procedure to the data in bins of \mXs\ and \mll.
We use the nominal Gaussian shape for all bins, as we found no significant
shape dependence on \mXs\ or \mll.

We evaluate the systematic uncertainties in the signal yield
by varying the signal Gaussian parameters (mean and width)
and the signal shape (using asymmetric signal shapes) within the 
constraints allowed by the charmonium-veto sample.
The amount of charmonium and hadronic peaking background
is varied,
as well as the shape of the latter.

Uncertainties in the signal efficiency originate from
the detector modeling and from the simulation of signal decays.
From control data samples we find uncertainties of
1.3\% per track (0.8\% per $\pi^+$) in the tracking efficiency;
0.85\% per electron, 1.55\% per muon,
1.0\% per $K^+$, and 1.4\% per $\pi^+$ in the
charged-particle identification efficiency;
1.5\% per \KS\ and 5.6\% per $\pi^0$ in the reconstruction
efficiency.
We check the efficiency of the cut in ${\cal L}_R$
with the charmonium-veto sample and take the discrepancy
with the simulation (3.7\%) as the uncertainty.
The fraction of signal cross feed included in the signal Gaussian is
varied by $\pm 100\%$.
Altogether, the uncertainty from detector modeling is 8.3\%.

The dominant source of uncertainty (18\%)
arises from the decay model.
The fractions of \BtoKll\ and \BtoKstarll\ decays
are both increased (or decreased) together
by the theoretical uncertainties~\cite{Ali02},
resulting in the largest contribution (16\%).
Parameters of the Fermi motion model are varied within limits
allowed by measurements of hadronic moments in semileptonic
$B$ decays~\cite{CLEO_moments}
and the photon spectrum in inclusive $B \to X_s\:\gamma$
decays~\cite{CLEO_bsgamma}.
The transition point (in \mXs) between the \BtoKstarll\
and \btosll\ decay models
is varied by $\pm 0.1$~\gevcc.

Hadronization uncertainties affecting the region
$\mXs > 1.1$~\gevcc total 4.9\% and are evaluated as follows.
The ratio between the generator yield for decay modes containing
a \KS\ and that for modes containing a charged kaon
is varied within the range $0.50 \pm 0.05$,
to allow for isospin violation.
Similarly, the ratio between modes with one and no $\pi^0$
is varied within the range $1.0 \pm 0.5$,
and the ratio between two-body and three-body hadronic
systems is varied within the range $0.5 \pm 0.3$.
Uncertainties in the last two ratios are set by the
level of discrepancy between data and simulation
as measured in~\cite{BaBar_bsgamma}.
The uncertainty in the fraction of modes
with pion or kaon multiplicities different from those
used in this analysis,
or with photons
that do not originate from $\pi^0$ decays but rather
from $\eta$, $\eta^\prime$, $\omega$, etc,
is estimated by varying
these different fractions by $\pm 50\%$.

Table~\ref{tab_results} summarizes the results of the analysis.
For the combined \BtoXsll\ branching fraction we assume
equal branching fractions in the electron and muon channels,
and average over both channels.
Table~\ref{tab_bins} shows the partial branching fractions in
several dilepton and hadronic mass ranges.

\begin{table}
 \caption{Partial branching fractions in bins of dilepton
  and hadronic mass averaged over electrons and muons.
  The first error is statistical and the second error is systematic.
  }
 \baselineskip=28pt
 \begin{center}
 \begin{tabular}{cccc}
  \hline \hline 
  $\mll$             & ${\cal B}$            & $\mXs$     & ${\cal B}$ \\
  (\gevcc)           & $(10^{-6})$           & (\gevcc)   & $(10^{-6})$\\
  \hline\hline
  0.2 -- 1.0         & $0.08 \pm 0.36^{+0.07}_{-0.04}$ & 0.4 -- 0.6 & $0.53 \pm 0.17 \pm 0.04$ \\
  1.0 -- 2.0         & $1.6 \pm 0.6 \pm 0.5$ & 0.6 -- 0.8 & $0.32 \pm 0.24 \pm 0.04$ \\
  2.0 -- $m_{J/\psi}$& $1.8 \pm 0.8 \pm 0.4$ & 0.8 -- 1.0 & $0.64 \pm 0.40 \pm 0.08$ \\
  $m_{J/\psi}$ -- $m_{\psi^\prime}$
                     & $1.0 \pm 0.8 \pm 0.2$ & 1.0 -- 1.8 & $3.5 \pm 2.1 ^{+1.1}_{-0.9}$ \\
  $m_{\psi^\prime}$ -- 5.0
                     & $0.64 \pm 0.32^{+0.12}_{-0.09}$ &   &  \\ \hline
  1.0 -- 2.45        & $1.8 \pm 0.7 \pm 0.5$           &   &  \\
  3.8 -- 5.0         & $0.50 \pm 0.25^{+0.08}_{-0.07}$ &   &  \\
  \hline
 \end{tabular}
  \label{tab_bins}
 \end{center}
\end{table}

  We search for \CP\ violation in the \BtoXsll\ decay by
performing separate fits to $B$ and $\overline{B}$ final
states, where the final state flavor is determined by the
kaon and pion charges (modes with $X_s$ = \KS, $\KS\pi^0$,
or $\KS\pi^+\pi^-$
are not used).
We find $N_{sig}^{\overline{B}} = 14.7 \pm 6.5({\rm stat})$ and
$N_{sig}^B = 22.9 \pm 7.4({\rm stat})$,
corresponding to an asymmetry
$A_{\CP} \equiv (N^{\Bb} - N^{\B}) / (N^{\Bb} + N^{\B})
        = -0.22 \pm 0.26({\rm stat})$.
The systematic uncertainty
is taken to be the statistical uncertainty in the asymmetry
measured in the charmonium-veto sample
$A_{\CP}^{c\bar{c}s} = -0.005 \pm 0.016({\rm stat})$.


In summary,
we observe a signal of $40 \pm 10({\rm stat}) \pm 2({\rm syst})$
\BtoXsll\ events
with a statistical significance of $4.3\:\sigma$.
The corresponding branching fraction
$  {\cal B}(B \to X_{s}\:\ell^{+} \ell^{-}) = 
 \left(5.6 \pm 1.5({\rm stat}) \pm 0.6({\rm exp~syst})
 \pm 1.1({\rm model~syst})\right) \times 10^{-6}$
for $\mll > 0.2$~\gevcc
agrees well with predictions~\cite{Ali02}
and a previous measurement~\cite{Belle_sll}.
In restricted dilepton mass ranges below and above the charmonium
region (see bottom of Table~\ref{tab_bins}),
the partial branching fractions also agree
with predictions~\cite{GHIY03}.
At low hadronic mass, the partial branching fractions
are consistent with existing $B \to K^{(\ast)} \ell^+ \ell^-$
measurements~\cite{Belle_Kll,BaBar_Kll}.
We determine the direct \CP\ asymmetry to be
$A_{\CP} = -0.22 \pm 0.26({\rm stat}) \pm 0.02({\rm syst})$,
in agreement with a predicted asymmetry of
$(0.19^{+0.17}_{-0.19})\times 10^{-2}$ in the SM~\cite{AH99}.
Overall, we find no evidence for contributions
from physics beyond the SM.

We thank Gudrun Hiller and Tobias Hurth
for helpful discussions.
We are grateful for the excellent luminosity and machine conditions
provided by our \pep2\ colleagues, 
and for the substantial dedicated effort from
the computing organizations that support \babar.
The collaborating institutions wish to thank 
SLAC for its support and kind hospitality. 
This work is supported by
DOE
and NSF (USA),
NSERC (Canada),
IHEP (China),
CEA and
CNRS-IN2P3
(France),
BMBF and DFG
(Germany),
INFN (Italy),
FOM (The Netherlands),
NFR (Norway),
MIST (Russia), and
PPARC (United Kingdom). 
Individuals have received support from the 
A.~P.~Sloan Foundation, 
Research Corporation,
and Alexander von Humboldt Foundation.

\end{document}